\documentclass[12pt]{book}

\usepackage[dvips]{graphicx,color}
\usepackage{makeidx,tsukuba}

\makeauthorindex
\makeindex

\begin{document}

\BookTitle{\itshape The 28th International Cosmic Ray Conference}
\CopyRight{\copyright 2003 by Universal Academy Press, Inc.}
\pagenumbering{arabic}

\chapter{
Cosmic Ray Propagation and Acceleration}

\author{Igor V.~Moskalenko\\
{\it NASA/Goddard Space Flight Center, Code 661, Greenbelt, MD 20771, USA\\
JCA/University of Maryland, Baltimore County, Baltimore, MD 21250, USA
}
}

\section*{Abstract}
Theoretical views on particle acceleration in astrophysical sources
and propagation of cosmic rays (CR) depend very much on the quality of
the data, which become increasingly accurate each year and therefore
more constraining.  On the other hand, direct measurements of CR are
possible in only one location on the outskirts of the Milky Way and
present only a snapshot of very dynamic processes.  The theoretical
papers presented during the conference offer exciting insights into
the physics of cosmic accelerators and processes which underlie the
measured abundances and spectra of CR species.

This paper is based on a rapporteur talk given at the 28th
International Cosmic Ray Conference held on July 31--August 7, 2003 at
Tsukuba. It covers the sessions OG 1.3 Cosmic ray propagation, OG 1.4
Acceleration of cosmic rays, and a part of HE 1.2 Theory and
simulations (including origins of the knee).

\section{Introduction}

The reviewed material can be organized as follows. OG 1.3 session (22
papers) covers production and propagation of light secondary isotopes,
antiprotons and antinuclei, TeV electrons in CR and their sources, new
effects in CR propagation, and new propagation models and codes. OG
1.4 session (15 papers) is mainly devoted to particle acceleration in
non-relativistic shocks in supernova remnants (SNRs), Galactic winds,
and galaxy clusters, and particle acceleration in relativistic shocks,
jets, and pair plasmas. The reviewed part of HE 1.2 session (8 papers)
discusses the origin of the ``knee'' as the result of diffusion or
acceleration by \emph{Galactic} sources, while the rest of the session
is covered in the rapporteur talk by Takita [23].
Although the number of papers
to review is relatively large, the subject, which can be broadly
described as ``theoretical astrophysics'', dictates an individual (not
statistical) approach since every contribution is different. A
complete and comprehensive discussion of all the results presented
during the conference, however, is not attempted here due to  space
limitations. I thus offer my apologies to the authors who feel their
work is not given sufficient coverage. The preference will be given
to new ideas and results interesting to the CR community, while the
choice necessarily reflects the Rapporteur's personal view of the
subject. This delicate matter is complicated by my own authorship of
five contributed papers in session OG 1.3; those papers I attempt
to appraise without a personal bias. The citation scheme adopted here
identifies contribution papers by the first author name, session, and
the page number.

\section{Propagation of Low-Energy Cosmic Ray Nuclei}

Acceleration of particles in the sources and their propagation in the
interstellar medium are traditionally separated subjects, although
there is no clear division between the ``sources'' and the
``interstellar medium''. Under  sources people usually intend SNRs,
whose energetics can sustain the observed CR density, but the wider
context includes also pulsars, stellar winds, and ensembles of shocks
in superbubbles. Particles escaping the acceleration sites
continuously (pulsars, winds, superbubbles) or on a short time scale
(SNRs) are injected into the interstellar medium (ISM) and become what
we call CR. Because of the huge Galactic volume, the CR particles
remain contained in the ISM for some 10 Myr before escaping into
intergalactic space. Energy losses and stochastic re-acceleration by
magnetic turbulence (2nd order Fermi acceleration) during the
propagation as well as escape from the Galaxy change the initial
spectrum of particles. The destruction of primary nuclei via
spallation gives rise to secondary nuclei and isotopes which are rare
in nature, antiprotons, and pions which decay producing $\gamma$-rays
and secondary positrons and electrons.

The wealth of information contained in the CR isotopic abundances makes
it possible to study various aspects of their acceleration and
propagation in the interstellar medium as well as the source
composition. Stable secondary nuclei tell us about the diffusion
coefficient and Galactic winds (convection) and/or re-acceleration in
the interstellar medium.  Long-lived radioactive secondaries provide
constraints on global Galactic properties such as, e.g., the Galactic halo
size. Abundances of K-capture isotopes, 
which decay via electron K-capture after attaching an electron from
the ISM, can be used to
probe the gas density and acceleration time scale.  Details of the
Galactic structure, such as the non-uniform gas distribution (spiral
arms, the Local Bubble), radiation field and magnetic field (regular
and random) distributions, and close SNRs may also affect the local
fluxes of CR particles. Heliospheric influence (modulation) changes
the spectra of CR particles below $\sim10-20$ GeV/nucleon as they
propagate from the boundaries of the solar system  toward the
orbits of the inner planets; these distorted spectra are finally
measured by balloon-borne and satellite instruments.  Current
heliospheric modulation models are based on the solution of Parker's
transport equation [18] and include four major processes:  convection,
diffusion, drifts, and adiabatic energy losses.  Individually, these
processes are well studied, however, their combined effect, which
produces the modulation, is still not fully understood.  The most
frequently used ``force-field'' approximation [7] with one parameter,
the modulation potential, accounts only for the effect of adiabatic
losses. It has been recently realized that direct information about
the fluxes and spectra of CR in distant locations is provided  by the Galactic 
diffuse $\gamma$-rays, therefore, complementing the local CR studies,
but this connection requires extensive modeling and  is yet to be
explored in detail.

Most efforts in OG 1.3 session are traditionally devoted to
interpretation of the data in the few hundred MeV to GeV energy range,
such as abundances of radioactive isotopes, light elements, spectrum
of antiprotons, and predictions of antinuclei and exotic particles in
CR.

\subsection{Radioactive Secondaries} \label{radioactive}
A conventional way to derive the propagation parameters is to fit the
secondary/primary ratio, e.g., B/C. While taken alone the stable
secondary/primary ratio does not allow one to derive a unique set of
propagation parameters; the radioactive isotope abundances, e.g.,
$^{10}$Be/$^9$Be ratio, are used to break the degeneracy. The
propagation parameters can be derived in a semi-phenomenological
approach, similar to what was used for decades in the Leaky-Box model,
by adjusting the energy dependence of the diffusion coefficient
\emph{ad hoc} to fit the B/C ratio, or by using a proper diffusion
model (including, e.g., diffusive reacceleration, convection) with
corresponding fitting of free parameters to match the B/C ratio. While
the formal procedure is quite clear, the nuclear production cross
section errors introduce considerable uncertainty into the propagation
parameters.  Hundreds of isotopes are involved in the calculation of
nuclear fragmentation and transformation of energetic nuclei in the
course of their interaction with interstellar gas, however, the widely
used semi-phenomenological systematics have typical uncertainties of
the order of 20\%, and can sometimes be wrong by a significant factor.

A steady state propagation model with cylindrical symmetry was used by
Molnar \& Simon (OG 1.3, p.1937) to study the abundance of $^{10}$Be
in CR, where the rigidity dependence of $D(R)/H$ (the ratio of the
diffusion coefficient to the halo size, $R$ being the rigidity) is
derived \emph{ad hoc} from the fitting to the B/C data. The low-energy
data $<150$ MeV/nucleon from Voyager, Ulysses, and ACE (to name only
recent experiments) on $^{10}$Be/$^9$Be ratio are all consistent and
require the halo size $H\sim4$ kpc, while the ISOMAX-98 data at
energies around 1 GeV/nucleon lie higher than predicted by models. The
authors offer two explanations to solve the discrepancy: cross section
errors and the energy dependence of the halo size. The reaction
$^{11}$B+p$\to ^{10}$Be is one of the most important channels of
$^{10}$Be production, but four available data points hardly allow for
any meaningful conclusion on the cross section; a change in the cross
section from 5 to 25 mb above 1 GeV/nucleon consistent with the cross
section error bars improves the agreement with the ISOMAX data. The
halo size also affects the $^{10}$Be/$^9$Be ratio. In particular, the
calculated ratio increases as the halo size decreases, an effect
connected with the decrease of the effective collecting volume of CR
(so that the produced radioactive species have less time to decay). The
halo size, which decreases with energy, can thus improve the
model. The interpretation of the latter effect is not, however,
straightforward. On one hand, as the energy of the particles increase,
they are less scattered by the magnetic turbulence in the halo and
should escape faster to  intergalactic space. On the other hand, in
a model with a Galactic wind, the halo size should increase with
energy (V\"olk, in the discussion) as low-energy particles are
convected easier and thus never return, while high energy
particles may still return. More accurate data from PAMELA and AMS
will provide a conclusive test.

Papers by Donato et al.\ (OG 1.3, p.1953) and Farahat et al.\ (OG 1.3,
p.1957) discuss the effect of the Local Bubble (LB) on the propagation
of radioactive isotopes. The LB is a low density region around the
sun, filled with hot H\,{\sc i} gas (e.g., [21]). The size of the
region is about 200 pc, and it is likely that it was produced in a
series of supernova (SN) explosions. While this relatively small
structure may produce only minor effects on stable nuclei, the
radioactive isotopes may be underproduced in this region. Four
radioactive isotopes, $^{10}$Be, $^{26}$Al, $^{36}$Cl, and $^{54}$Mn,
are commonly used to probe the effective Galactic volume filled with
CR and derive the confinement time of CR in the Galaxy. Their
half-lives range from $3.07\times10^5$ yr ($^{36}$Cl) to
$1.60\times10^6$ yr ($^{10}$Be) with the shortest half-life being most
sensitive to the local structure. The half-life of
$^{54}$Mn($\beta^-$) is the only one of the four which is not measured
directly and remains the most uncertain.  Assuming that the diffusion
coefficient in the LB is equal to the Galactic average, the effect is
easy to estimate: the calculated propagation lengths of these
radioactive  species appear to be comparable to the LB size.
Inclusion of the LB into a model thus should lower the halo size
compared to calculation without the LB.  This simple picture is
undermined by the fact that the propagation parameters (and the
diffusion coefficient) in the LB are unknown while those derived from
the B/C ratio give an \emph{average} diffusion coefficient, which
samples a large Galactic volume. A semi-analytical propagation model
(with reacceleration \emph{and} convection)
and a formal $\chi^2$ fit to the B/C, $^{10}$Be/$^9$Be, and
$^{36}$Cl/Cl ratios were used by Donato et al., who found that taken
together the ratios are compatible with ACE data if the LB hole radius
is $r_{\rm hole}\sim60-100$ pc, while $r_{\rm hole}=0$ is
disfavored. The diffusion coefficient index treated as a free
parameter was found to be $\delta=0.5-0.8$. The latter is apparently
larger than in conventional models with reacceleration, and besides,
explaining the CR anisotropy measurements at high energies (see
Fig.~\ref{fig:ptuskin}) may be problematic. The $^{26}$Al/$^{27}$Al
ratio calculated in the same model was found to be inconsistent with
the ACE data and marginally consistent with Ulysses data, which 
the authors conclude may indicate some errors in the data 
(nuclear or astrophysical) used in
the analysis. Indeed, the cross section errors are often the reason
for inconsistency in propagation models. For Al isotopes the main
progenitor is $^{28}$Si, while the contribution of $^{27}$Al to $^{26}$Al
is also important. For isotopes of Cl the main progenitor is
$^{56}$Fe, but the contribution of many lighter nuclei is equally
important. Comparison with the data for those reaction channels shows
that the semi-phenomenological approximations produce bad fits (e.g., [15,24]).
A difference in the calculated $^{26}$Al/$^{27}$Al and
$^{36}$Cl/$^{35}$Cl ratios at low energies in models with/without the
LB has been also reported by Farahat et al., who used a Markov
stochastic method to calculate the path-length probability
distribution, followed by a weighted slab calculation.

\subsection{Light Stable Secondary Isotopes} \label{stable}
Isotopes of light elements in CR are almost all secondaries 
produced in spallations of heavier nuclei on interstellar gas
Galaxy-wide, and their abundances reflect the propagation history of
CR; an example is the best-measured B/C ratio, which is used to fix
propagation parameters. On the other hand, weak local sources, e.g., old SNRs
in the LB, may still accelerate particles out of the ISM by weak
shocks [4] adding to the Galactic CR at low energy. This may change
the local CR composition of progenitor nuclei adding a fresh local
``unprocessed'' component. Deviations of the calculated
abundances of secondaries from measurements thus can tell us about our
local environment. A study of the abundances of light elements, $^2$H,
$^3$He, Li, Be, B, in reacceleration models with/without the LB has been
reported by Moskalenko et al.\ (OG 1.3, p.1917), who use the GALPROP
propagation code. Apparently both reacceleration models with/without
the LB are consistent with the ACE data given the large error bars.
$^3$He/$^4$He ratio agree well with IMAX92 data indicating that $^3$He
in CR is mostly secondary. The He/Si ratio in the LB sources required to
match the $^3$He/$^4$He ratio is 2 times that of the Galactic CR
sources indicating that the material was possibly diluted in the LB
medium before the acceleration. Presented were also calculations of the
$^2$H/$^4$He ratio, which is a factor of 1.5 less than the IMAX92
value, but this may be connected with systematic errors in the data.

The deuteron flux as evaluated in simplified propagation models is
presented by Sina et al.\ (OG 1.3, p.1973). The models (Leaky-Box,
convection with a constant wind speed, and stochastic reacceleration)
indicate somewhat different behavior of the flux below 1
GeV/nucleon. The calculations are consistent within a factor of 2 or
less with the BESS data taken in a series of flights during different
periods of solar activity. The reacceleration model generally
agrees better with the data 
predicting the deuteron flux (modulation potential $\Phi=500$ MV) just
above the BESS-97 solar minimum data. Though, the modulation level can
be deduced from the CR data and neutron monitors only approximately,
the increased levels of modulation will result in a decrease of deuteron
flux consistent with the general trend, while other models would underestimate
the deuteron flux.

Stable CR species probe a large region of the Galaxy. The size
of this region depends essentially  on the nuclear disintegration cross
section and is smaller for heavier nuclei. A regular Galactic magnetic
field may cause preferential propagation in azimuthal direction making
some sources more important even if they are more distant. The global
structure of Galactic magnetic field is currently derived from
observations of rotation measures of more than 500 pulsars [8]. It is
best described by two distinct components, a bi-symmetric spiral field
in the disk with reversed direction from arm to arm, and an azimuthal
field in the halo with reversed directions below and above the
Galactic plane. Codino \& Plouin (OG 1.3, p.1977) present their
simulations of the propagation of CR particles from an acceleration
site to the solar system. They show that the sources contributing to
the local CR flux are predominantly located along the principal
magnetic field line. Contrary to the Leaky-Box assumption, the mean
trajectory length of CR particles reaching the solar system decreases
with the increase of the atomic number and reaches a constant level of
about 150 kpc for $A>30$ at 10 GeV/nucleon.

Errors in the nuclear production cross sections are frequently quoted
by many authors as one of the main reasons for uncertainty in
propagation parameters, which are usually derived using B/C and
$^{10}$Be/$^9$Be ratios. This clearly becomes a factor hindering
further progress since the accuracy of the cross sections used in
astrophysics is far behind the accuracy of recent CR measurements.
The scarce cross section data alone cannot be used to produce a reliable
evaluation of the cross sections, while current nuclear codes and
semi-empirical parameterizations also fall short of predicting cross
section behavior for the whole range of target nuclei and incident
energies. An auxiliary paper by Moskalenko \& Mashnik (OG 1.3, p.1969)
uses a collection of cross section data from LANL nuclear database to
evaluate the most important channels of production of Li, Be, and B
isotopes. The individual cross section evaluations are tested against
isobaric and cumulative data where available. Such work should help to
eventually reduce the discrepancy in the results of different groups.

\subsection{Secondary Antinuclei}

The baryonic asymmetry of the universe is one of the most intriguing puzzles.
Antinuclei in CR may present an opportunity to test the
baryonic asymmetry or uncover evidence for 
annihilation of the non-baryonic Dark Matter particles (weakly
interacting massive particles -- WIMPs) whose existence is predicted
by supersymmetric models [12]. The absence of baryonic antimatter in
the universe or its presence in equal amounts with matter in
spatially separated domains, if found, would have a dramatic impact on
theories of grand unification, Big Bang nucleosynthesis, and
cosmology.  The lower limit to the domain size perhaps corresponds to
the scale of  a galaxy cluster, of the order of 20 Mpc, because the
intercluster voids can prevent annihilation. Due to the very low
probability of formation of \emph{secondary} antinuclei in CR, the
detection of an antinucleus with $Z<-1$ in CR would be a ``smoking
gun'' of  new physics. The production of secondary  antinuclei $\bar
d$, $\bar t$, $^3$\rlap{\hbox{\lower-7pt\hbox{---}}}\hbox{He},
$^4$\rlap{\hbox{\lower-7pt\hbox{---}}}\hbox{He} in the atmosphere and
in CR interactions with interstellar gas has been evaluated by Baret
et al.\ (OG 1.3, p.1961), who used a coalescence model.  The
coalescence model is based on the hypothesis that nucleons produced in
collisions of energetic particles fuse into light nuclei whenever the
momentum of their relative motion is smaller than the coalescence
radius in  momentum space. Galactic propagation has been evaluated
using the Leaky-Box and diffusion models. The atmospheric background
of antinuclei is shown to be considerably less than the secondary
antinuclei flux in CR thus does not contribute much to the experimental
uncertainty in searches for antimatter nuclei in CR.  In the search for 
primordial antimatter $^4$\rlap{\hbox{\lower-7pt\hbox{---}}}\hbox{He}
is a good candidate because of the vanishing background of secondary
particles.

The indirect searches for exotic sources such as WIMPs or primordial
black holes also concentrate on signals in CR with low background. The
most often discussed are searches for spectral signatures in
antiprotons, positrons, and diffuse $\gamma$-rays. Secondary
antiprotons and positrons (and diffuse $\gamma$-rays) are products of
interactions of mostly CR protons and helium nuclei with interstellar
gas. Due to the kinematics of $pp$-interactions, the spectrum of
antiprotons peaks at about 2 GeV decreasing sharply toward lower
energies, a unique shape distinguishing it from other CR species. If
the CR propagation model and heliospheric modulation is correct, a
model that describes any secondary to primary ratio should equally
well describe all the others: B/C, sub-Fe/Fe, $\bar p/p$ ratios, as
well as spectra of nuclei, positrons, and diffuse Galactic continuum
$\gamma$-rays. It appears that relatively simple propagation models
can not account simultaneously for the nuclei component in CR and
antiprotons. In particular, a ``standard'' reacceleration model pose a
problem since it produces too few  antiprotons\footnote{For an
alternative viewpoint see [5].}  [14,16] though it works well for
other CR species.

Moskalenko et al.\ (OG 1.3, p.1921) discuss the reasons for the
deficit of antiprotons as calculated in the reacceleration model.  The
uncertainties are described and divided into four categories:
propagation models and parameters, production cross sections of
isotopes and antiprotons, heliospheric modulation, and systematic
errors of measurements. In the first category, the most uncertain are
the spectra of CR species in the distant regions of the Galaxy; a
direct test is provided by diffuse $\gamma$-rays, but the message of
the ``GeV excess'' [10] is not fully understood yet. Nuclear
production cross section errors are one of the major concerns; while
there are not enough measurements, semi-empirical systematics are often
wrong by 20\% and sometimes by a considerable factor (for more details
see Sections \ref{radioactive} and \ref{stable}). The errors
introduced by heliospheric modulation and instrumental errors are
difficult to account for, but their effect could be reduced by careful
choice of data. If, however, interstellar propagation and heliospheric
modulation are correct, the discrepancy may indicate new phenomena.
The paper discusses new ideas such as the change in the character of
the diffusion at low energies, the effect of the local environment (e.g.,
fresh low-energy CR nuclei from the Local Bubble), and a population of
low-energy proton sources, etc. New accurate measurements of
antiprotons and CR at both low and high energies can help to resolve
the issue.

\subsection{New Approaches to CR propagation}

Galactic CR are an essential factor determining the dynamics and
processes in the ISM. The energy density of relativistic particles is
about 1 eV cm$^{-3}$ and is comparable to the energy density of the
interstellar radiation field, magnetic field, and turbulent motions of
the interstellar gas. While propagation of CR is often considered as
propagation of test particles in given magnetic fields, the stochastic
acceleration of CR by MHD waves causes damping of waves on a small
scale, which, in turn, affects the propagation of CR. This illustrates
a need for a self-consistent approach employed by Ptuskin et al.\ (OG
1.3, p.1929) who develop a formalism for the dissipation of
hydromagnetic waves on energetic particles for the cases of Kolmogorov
$W(k)\propto k^{-5/3}$ and Iroshnikov-Kraichnan $W(k)\propto k^{-3/2}$
spectra of the turbulences. In this formalism the mean free path of CR
particles depends not only on the spectrum of turbulences, but also on
the CR (mostly protons) momentum distribution function itself. The
Kolmogorov type cascade is not very much affected by the damping. The
Iroshnikov-Kraichnan type cascade is significantly affected; this
should lead to a modification of  CR transport below $\sim$10
GeV/nucleon. In particular, the diffusion coefficient will increase
toward  low energies, an effect used \emph{ad hoc} in the
Leaky-Box model. This qualitative analysis is confirmed by numerical
calculations by Moskalenko et al.\ (OG 1.3, p.1925), who employed the
GALPROP propagation code and used an iterative numerical procedure
starting from an undisturbed diffusion coefficient. It offers a new
explanation of the peak in the secondary/primary ratio at a few
GeV/nucleon:  because of the damping, the amplitude of short waves is
small and thus the low energy particles rapidly escape the Galaxy
without producing secondaries. The preliminary analysis shows that
inclusion of the damping allows one to reproduce the B/C ratio and
obtain a good agreement with spectra of CR nucleons using a unique
power-law injection spectrum. Another interpretation, that the peak is
produced by CR reacceleration [20] on an undisturbed Kolmogorov-type
spectrum, remains as a viable alternative.

\subsection{Propagation Models and Codes}

Though the main features of CR propagation in the Galaxy seem to be
well established [3], some simplifications and inconsistencies
remain. The usual approach is to impose Galactic boundaries beyond which
particles escape freely. Hareyama et al.\ (OG 1.3, p.1941, p.1945) and
Shibata et al.\ (OG 1.3, p.1949) develop an analytical model of a
boundary-less Galaxy, where the boundaries are replaced with
exponential forms of the diffusion coefficient, gas density, and
sources, both in $R$ and $z$. The particle energy change (energy
losses, reacceleration, convection) is neglected. The spectra of
primaries and B/C and (Sc+Ti+V)/Fe ratios are reproduced above 5
GeV/nucleon given a set of fitting parameters; $^{10}$Be/$^9$Be ratio
is also calculated.

While analytical methods explore new approaches, numerical methods
offer realistic treatment of CR propagation and calculation of many CR
species simultaneously in the same model. Such approach takes
advantage of independent constraints from many different kinds of
data.  The choice and optimization of a numerical scheme is the first
necessary step in the development of a new propagation code. Busching
et al.\ (OG 1.3, p.1981, p.1985) test a modified
leapfrog/DuFort-Frankel scheme. A simplified time-dependent transport
equation (cylindrical symmetry) has been expanded using Bessel
functions in $r$ and trigonometrical functions in $\varphi$, where the
particle momentum is treated as a parameter. This reduces it to a
system of time dependent 1-D equations for the expansion coefficients,
which is solved numerically. The authors present simulations of the
intensity variation of CR density (at 10 GeV/nucleon) during a period
of 10 Myr in a cell of 500 pc $\times$ 500 pc containing a SNR.

GALPROP is a well-developed realistic numerical model of CR
propagation, which is used in numerous applications. The source code
(C++) is regularly updated and posted on the Web together with output
FITS files (models). However, reading the FITS format requires some
programming and may apparently be not convenient for everyone.
Moskalenko et al.\ (OG 1.3, p.1925) are developing a Web-based interface
for the GALPROP models that supports various formats. The interface will
include a simple form in the Web browser while the output results will
be provided as computer readable tables and graphics files. For every
posted model a user may request any of the following: spectra of any
combination of isotopes or elements $Z\leq28$, arbitrary isotopic
ratios vs.\ kinetic energy, isotopic distribution for a given $Z$,
elemental and isotopic abundance at arbitrary energy, as well as
antiprotons, electrons and positrons. First, it will be available for
the solar vicinity and later for the whole Galaxy.

\section{Local Sources and TeV Electrons}

\begin{figure}[tb]
\centerline{
   \includegraphics[width=0.5\textwidth,height=0.47\textwidth]{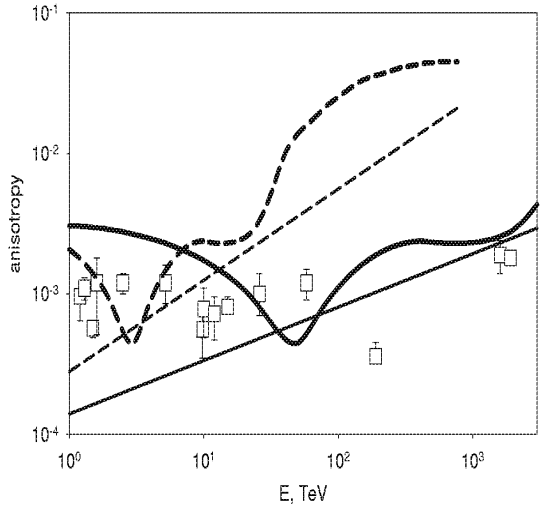}}
   \vspace{-0.5pc}
\caption{The CR anisotropy produced by local supernovae (thick curves)
and the expected fluctuation anisotropy (thin straight lines) in the
reacceleration (solid curves)  and the plain diffusion (dashed curves)
models (Ptuskin et al., OG 1.3, p.1933).
\label{fig:ptuskin}}
\end{figure}

In studies of cosmic-ray propagation and diffuse continuum
$\gamma$-ray emission from the Galaxy it has usually been assumed that
the source function can be taken as smooth and time-independent, an
approximation justified by the long residence time ($>10^7$ years) of
cosmic rays in the Galaxy. However, the inhomogeneities have
observable consequences, and their study is a step toward a
``realistic'' propagation model based on Galactic structure and
plausible source properties. The random nature of CR sources leads to
fluctuations of CR intensity in space and time. Ptuskin et al.\ (OG
1.3, p.1933) study the effect of nearby SNRs on the CR anisotropy at
1--1000 TeV where the energy losses of protons and heliospheric
modulation are not important. In the case of distributed sources, the
anisotropy calculated in the reacceleration model with the diffusion
coefficient $D\propto E^{0.3}$ agrees better with the data above
$\sim$10 TeV than in the plain diffusion model ($D\propto E^{0.54}$),
which is a standard result. Inclusion of nearby SNRs improves the
agreement of the data with the reacceleration model, where the most
important contributions  come from Vela and S 147
(Fig.~\ref{fig:ptuskin}). Inclusion of a very young and close SNR RX
J0852.0--4622 (0.2 kpc, 700 yr) would dramatically change the
predicted anisotropy; however it is probable that the source is still
in a free expansion stage with accelerating particles confined inside
the remnant.

\begin{figure}[tb]
    \includegraphics[width=0.5\textwidth,height=0.47\textwidth]{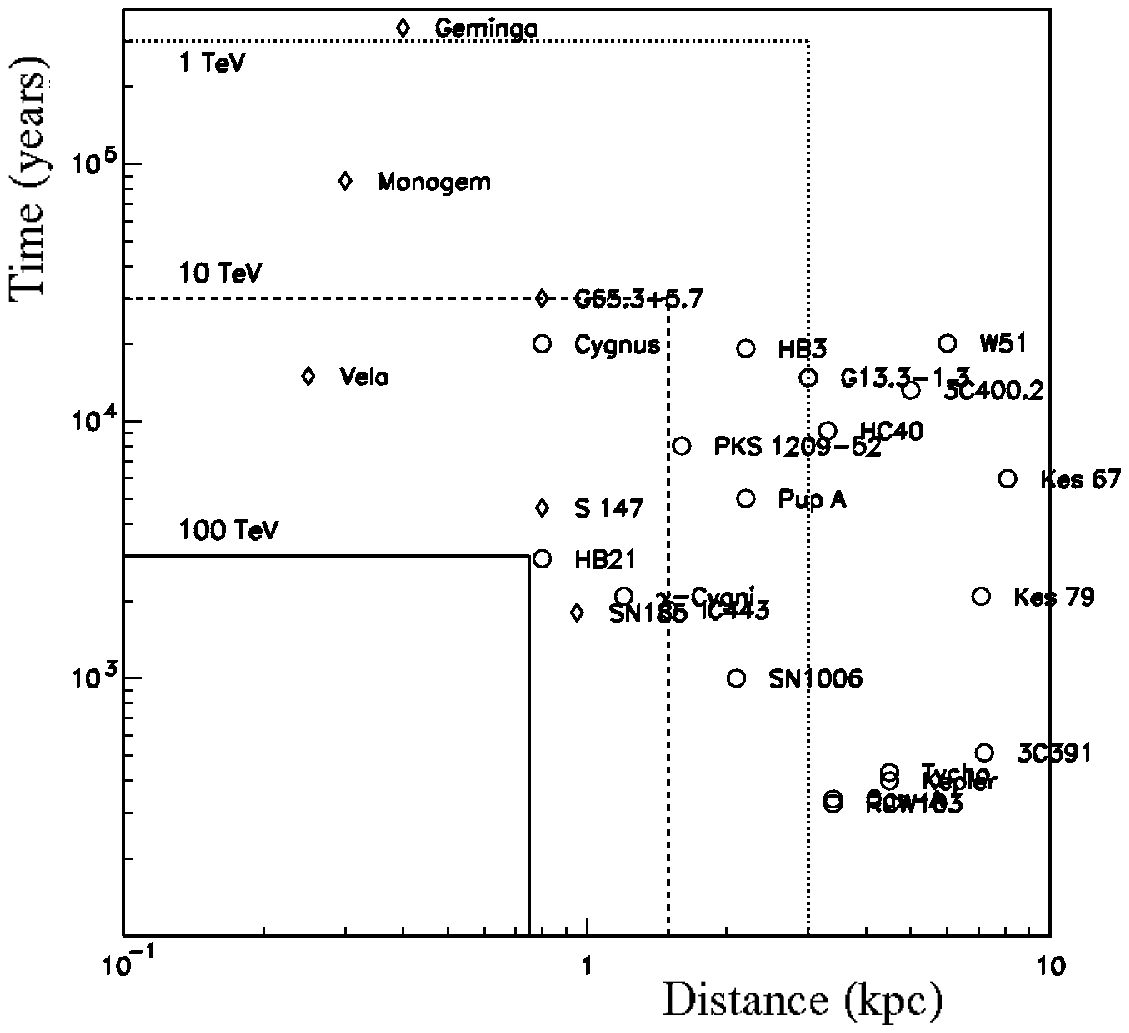}
\hfill
    \includegraphics[width=0.5\textwidth,height=0.45\textwidth]{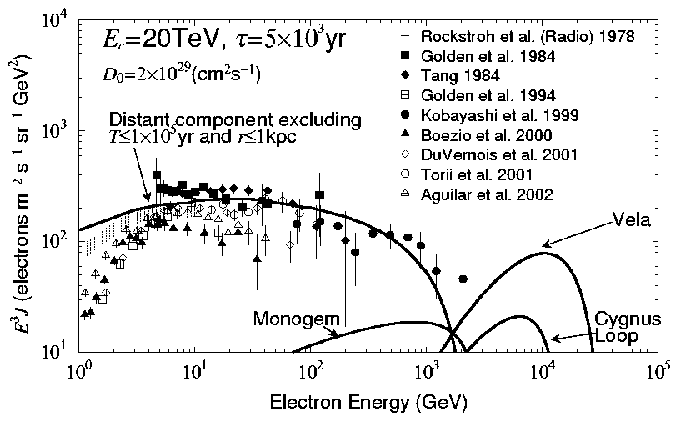}
    \vspace{-0.5pc} \begin{minipage}[tl]{0.48\textwidth}
\caption{Compilation of nearby shell-type and plerion SNR distances
and ages.  Line boxes represent limiting energies for electrons to
reach Earth (Swordy, OG 1.3, p.1989).
\label{fig:swordy}}
   \end{minipage}  
\hfill 
   \begin{minipage}[tl]{0.48\textwidth}
\caption{The local electron spectrum with  estimated contributions of
nearby sources. The source spectrum has a cut-off at 20 TeV, $\tau=5$
kyr (Yoshida et al., OG 1.3, p.1993).
\label{fig:yoshida}}
   \end{minipage}
\end{figure}

For electrons at very high energies, where the energy losses due to
inverse Compton and synchrotron emission are rapid, the effect of the
stochastic nature of CR sources becomes even more apparent.  Swordy
(OG 1.3, p.1989) and Yoshida et al.\ (OG 1.3, p.1993) discuss
fluctuations of TeV electrons due to nearby SNRs.  For the typical
energy density of Galactic radiation and magnetic fields of 1 eV
cm$^{-3}$, the energy loss timescale is $\sim3\times10^5$ yr at 1 TeV,
and becomes as short as $\sim3\times10^3$ yr at 100 TeV. A cutoff in
the electron spectrum at very high energies is thus unavoidable
because of both large energy losses and a discrete nature of the
sources. This is similar to the GZK effect for ultra high energy CR, but
it should be observable in the electron spectrum at much lower
energies (though a smaller flux of secondary electrons should be
present in CR at all energies). The analysis of nearby shell-type SNRs
has shown that the electron spectrum should have a cut off between 30
TeV and 100 TeV as measured near the solar system
(Fig.~\ref{fig:swordy}). Yoshida et al.\ continue earlier studies of
the propagation of very-high-energy electrons from local sources by
Nishimura and collaborators. Their calculations predict that some
nearby SNRs are possibly capable of producing unique identifiable
features in the CR electron spectrum at 1--30 TeV
(Fig.~\ref{fig:yoshida}), where the important parameters are the
distance and the age of a SNR. The most promising candidate sources of
TeV electrons are Vela, Cygnus Loop, and Monogem. Very-high-energy
electron measurements give a direct test of SNR origin of CR, but also
an important test of our local environment. The features in the
electron spectrum and the cutoff energy would immediately signal which
SNR(s) is/are affecting the local CR flux and to what degree with
implications for Galactic CR propagation models.

A study of two types of formalism for second-order Fermi
acceleration [9,20] in the interstellar medium is presented by Komori
(OG 1.3, p.1997). It shows that the approaches are almost identical in
the case of the same energy gain by a particle. The study concludes that
the consistency with the \emph{local} spectrum of CR electrons in
reacceleration models requires a break at $\sim10$ GeV in the electron
source spectrum. Since the local spectrum is affected by the solar
modulation, it is worth comparing with the spectral index of the
\emph{interstellar} electron spectrum. The latter can be deduced from
observations of synchrotron spectral index and intensity; the
synchrotron emission in 10 MHz -- 10 GHz band constrains the electron
spectrum in the $\sim1-10$ GeV range. Indeed, the low-frequency index
$\beta\hbox{\rlap{\hbox{\lower3pt\hbox{$\sim$}}}\lower-2pt\hbox{$<$}}2.5$
[19] indicates that the index of the injection spectrum must be below
$\sim2.0$. Casadei \& Bindi (OG 1.3, p.2001) discuss the local
spectrum of electrons between 0.8 GeV and 2 TeV. They use a collection
of electron CR measurements renormalized to the same value at 20
GeV. The fit shows that such spectrum can be represented by a power
law in kinetic energy with index 3.4. A model with a single source
also provides a good fit to the data above 10 GeV.

\section{Particle Acceleration in Shocks}

Shock acceleration is viewed as a ``universal'' acceleration mechanism
working well on different scales and in different astrophysical
environments [11].  The non-relativistic shocks are formed where the
pressure of a supersonic stream of gas drops to the much lower value of
the environment.  Examples of non-relativistic shocks include  the
Earth's bow shock, solar wind termination shock, shocks in expanding
SN shells, and galactic wind termination shocks, to name only a
few. The relativistic shocks give rise to  nonthermal particles
populations in astrophysical plasma known to exist in cores and jets
of active galactic nuclei and quasars, and probably existing in blast
waves of $\gamma$-ray bursts.  The early results were obtained in
the test-particle approximation, where the test particles do not modify
the shock. If acceleration is efficient, a  considerable amount of
energy is transferred to the accelerated particles and they  then
act on the shock itself, dynamically modifying it.  A significant flux
of energetic particles against cold material will generate plasma
turbulence. Escape of high-energy particles  from the shock will
allow the shock compression ratio to increase, which flattens the
spectrum of accelerated particles.  The non-linear effects are
important in SNR because of the strong shocks and long expansion time.
It is usually assumed that the Bohm diffusion coefficient is a good
approximation for strong shocks. Important parameters are 
the angle $\alpha$ between the shock normal and
the upstream magnetic field and $M_A=V/v_A$
Alv\'enic Mach number, where $V$ is the shock speed, and $v_A$ is the
Alfv\'en speed.  The characteristic feature of the shock
acceleration is the formation of a power-law spectrum of accelerated
particles in rigidity.

The papers in OG 1.4 section discuss particle acceleration in SNRs,
reacceleration of high-energy particles by shocks in the Galactic
wind, acceleration by shocks from mergers of galaxy clusters, shock
waves due to the formation of the large scale structure of the universe, and
relativistic shocks.

\subsection{Non-Relativistic Shocks}

The major sources of CR are believed to be SNe and SNRs with some
fraction coming from  pulsars, compact objects in close binary
systems, and stellar winds.  Observations of X-ray and $\gamma$-ray
emission from SNRs reveal the presence of energetic electrons thus
testifying to efficient acceleration processes near these objects,
while  evidence of accelerated hadrons is yet to be found.  The
radiation produced by an SNR depends on  the electron spectrum
\emph{inside} the SNR.  Lipari \& Morlino (OG 1.4, p.2027) calculate
the spectra of electrons and protons \emph{inside} young SNRs.  Their
solution for protons gives a standard spectrum with  power-law index
2. Inclusion of  energy losses for electrons due to  synchrotron
emission and inverse Compton scattering on cosmic microwave
background naturally produces a cut off in the electron spectrum.
They argue that there are two different populations of electrons,
those in the ``accelerator'' and those which are in the ``storage"
inside the remnant shell. The latter are particles trapped in
turbulent magnetic fields left behind the expanding shell.  The two
emission mechanisms, synchrotron and inverse Compton scattering,
produce two peaks in the photon spectrum. The relative intensity of
the peaks depends on the magnetic field strength inside the
remnant. The lower field results in a larger flux of TeV photons, but
also facilitates  particle escape, reducing the intensity of a
synchrotron peak.  Hoshino \& Shimada (OG 1.4, p.2047) use
particle-in-cell  simulations to study electron acceleration by  a
series of large amplitude electrostatic soliton-like  waves excited by
the Buneman instability in SNR shocks.

While SNRs are ``conventional'' sources of CR up to the knee energies
$\sim3\times10^{15}$ eV, the origin of higher energy  particles
remains unclear. Acceleration to higher energies may require new
classes of sources or multiple shocks (for more details see Section
\ref{sec:knee}).  V\"olk \& Zirakashvili (OG 1.4, p.2031) argue that
the knee must be a feature of the source spectrum. They propose CR
\emph{reacceleration} in a supersonic Galactic wind at distances  50
to 100 kpc from the plane as a mechanism to accelerate high-energy CR
particles beyond the knee energies, up to $Z\times10^{17}$ eV.  The
spiral compression of the wind formed due to the differential
Galactic rotation and the differences in the wind speed will form a
sawtooth series of forward shocks, which will re-accelerate
high-energy particles from the disk by $\sim$2 orders of
magnitude. Some fraction of these particles returns to the disk plane
giving rise to the component above the knee. The process is
reminiscent of the so-called Corrotating Interaction Regions in the
solar wind. There is no way in this mechanism to produce CR above
$10^{18}$ eV. The energy gain in this process is very slow so the
process of reacceleration may take some Myr.

The nonthermal radiation observed from some galaxy clusters implies
the presence of accelerated particles in the intracluster medium. The
acceleration by shocks formed due to  cluster mergers and its
implication for $\gamma$-ray emission from clusters is discussed in
two papers, Blasi \& Gabichi (OG 1.4, p.2051) and Takizawa et al.\ (OG
1.4, p.2059). Blasi \& Gabichi analysed the compression factors of the
shocks based on a model of mergers and formation of the hierarchical
structure of the universe. Their simulations indicate that particle
acceleration by shocks associated with mergers of clusters is
inefficient. The large majority of merger shocks have the Mach number
below 2 with correspondingly low acceleration capability; this
corresponds to particle spectra much steeper than required to
explain the observed nonthermal radio emission. In the case of shocks
formed due to the accretion in the gravitational field of the forming
cluster, Mach numbers can reach several hundred, which results in the
formation of a flat particle spectrum $\propto E^{-2}$. The
calculation of the $\gamma$-ray emission due to inverse Compton
scattering off the cosmic  microwave background photons shows that
future orbital $\gamma$-ray experiments, such as AGILE and GLAST, will
be able to detect a few to several dozens of clusters. The overall
contribution of clusters to the diffuse extragalactic $\gamma$-ray
emission (as derived from EGRET data [22]) is about 10\%, which is
significantly lower than some other estimates.  The cluster merger
shocks and random Alfv\'en waves in intracluster space are also
discussed by Takizawa et al., who used  N-body/hydrodynamic
simulations to study the electron spectral evolution due to
acceleration by the shocks.  The simulations show that accelerated
electrons are mostly confined to the shocks and exhibit their
morphology structure in radio and X-ray emission, while random
Alfv\'en waves may produce the radio halo.  Comparison with radio
emission from the Coma cluster indicates that the Alfv\'en wave power
spectrum may be significantly modified due to  particle-wave
interactions.

Three papers are devoted to simulation of CR acceleration in shocks
emerging in the large scale structure formation of the universe.
N-body/hydrodynamic simulations of formation of  large scale
structure by Ryu et al.\ (OG 1.4, p.2055) support scenarios in which
the intracluster medium contains significant populations of CR.
Cosmological shocks may be induced due to the accretion of baryonic
gas onto the nonlinear structures, such as sheets, filaments, knots,
subclump mergers, and chaotic flow motions.  The simulations show that
while  shocks with Mach numbers up to 100 are possible, the majority
of CR is accelerated by shocks with Mach number of a few. The
mean separation between shock surfaces is $\sim4h^{-1}$ Mpc at
present.  Kang \& Jones (OG 1.4, p.2039) study  CR injection and
acceleration efficiency in quasi-parallel cosmic shocks in 1-D
geometry for a wide range of Mach numbers and preshock conditions.
They find that a fraction of $10^{-4}-10^{-3}$ of the particles
which passed through the shock becomes CR. A new fast numerical scheme for
the time-dependent CR diffusion convection equation to study the
evolution of CR modified shocks is being developed by Jones \& Kang
(OG 1.4, p.2035).

\subsection{Relativistic Shocks}

Relativistic shocks are associated with the most energetic objects in
the universe, such as active galactic nuclei, quasars, and
$\gamma$-ray bursts. The particle velocity in such shocks is
comparable to the bulk velocity leading to substantial anisotropy of
the angular distribution of particles. In case it makes an angle to the
shock normal, the upstream magnetic field is strongly modified by the
Lorentz transformation to the downstream frame
and further modified by the shock compression.  The downstream
magnetic field is increased and tilted toward the plane of the shock
influencing both the shock jump conditions and particle acceleration.
The magnetic field intensity increases across the shock reflecting
particles and thus provides more effective energy gain.

Relativistic shock acceleration was the subject of four papers, by
Niemec \& Ostrowski (OG 1.4, p.2015, p.2019), by Virtanen \& Vainio
(OG 1.4, p.2023), and by Nishikawa et al.\ (OG 1.4, p.2063).  
The process of  first-order Fermi acceleration in the parallel and
oblique shocks in a pair plasma in the presence of finite-amplitude
magnetic field perturbations is discussed by Niemec \& Ostrowski. The
perturbations are assumed to have either a flat, $F(k)\sim k^{-1}$, or
a Kolmogorov spectrum, $\sim k^{-5/3}$. In the case of a parallel shock,
the long-wave finite-amplitude perturbations produce locally oblique
magnetic field configurations with the probability of particle
reflection depending on the turbulence amplitude. This leads to
shorter acceleration timescales. In the case of an oblique shock, the
exact shape of the spectrum depends on the amplitude of the magnetic
field perturbations and the wave power spectrum. Subluminal shocks
produce a spectrum that hardens below the cut-off energy, while
superluminal shocks lead to a spectrum steepening below the
cut-off. Electron acceleration in parallel shocks with finite
thickness is studied by Virtanen \& Vainio using test-particle
simulations, where the thickness of the shock is determined by the ion
dynamics. Electrons were injected in the downstream region. The finite
thickness of the shock reduces its efficiency, leading to a modest
acceleration resembling  adiabatic compression. An energy-independent
mean free path leads to a spectral index 3.2. The canonical value 2.2
is obtained in the case of a thin shock as a high energy limit for a mean
free path increasing with energy. Nishikawa et al.\ present a
simulation of particle acceleration and generation of the magnetic
field in relativistic jets using 3-D relativistic particle-in-cell
(REMP) simulations with and without  initial ambient magnetic fields.
They show that the Weibel instability is responsible for generating
and amplifying highly nonuniform, small-scale magnetic fields,  which
contribute to the electron's transverse deflection behind the jet
head, while accelerating particles parallel and perpendicular to the
jet propagation vector. The deflected electrons emit  ``jitter''
radiation which may help to understand the complex  time evolution
and/or spectral structure of $\gamma$-ray emission from astrophysical
objects.

\section{Origins of the Knee}\label{sec:knee}

The small change in the slope at $\sim3\times10^{15}$ eV of otherwise
almost featureless CR spectrum is known as the ``knee.''  Besides
numerous speculations, the origin of this feature  discovered more
than 40 years ago by a group of scientists at Moscow State University
still remains unexplained.  Because of the low flux of CR particles at
the knee energy,  direct measurements from balloons and spacecraft
are apparently inefficient;  indirect measurements using the
extensive air showers technique (Cherenkov light and  detector arrays)
is presently the only way to study CR with high statistics at energies
near or above the knee.  One of the main difficulties of the indirect
techniques is that the event reconstruction depends on hadronic
interactions at energies beyond those currently available in
accelerator experiments. This leads to one possible explanation that
the knee is the result of our incomplete understanding of the
properties of high energy particle interactions.  On the other hand,
the knee position is remarkably close to the highest possible energy
expected from the shock acceleration in SNRs.  Acceleration of
particles to energies beyond the knee still requires a satisfactory
explanation. The ideas discussed consider new classes of Galactic or
extragalactic sources, a local source, or acceleration by multiple
shocks.  Additionally, a drift of CR particles in the large scale
regular Galactic magnetic field,  Hall diffusion, may dominate the
regular diffusion at very high energies thus making it possible to
explain the knee  by the increased leakage of CR from the Galaxy.
Changes in the  amplitude and phase of the anisotropy and/or a
composition of CR at the knee region can perhaps provide necessary
clues to its origin, although the current uncertainties in the
composition are still large given the uncertainties in the underlying
hadronic interaction models.

The HE 1.2 papers are devoted to the origins of the knee in CR
spectrum and CR above the knee.

\subsection{Diffusion Mechanism}

The ``poly-gonato'' empirical model is proposed by Horandel et al.\
(HE 1.2, p.243), where the knee structure appears as the result of
many knees in the individual CR components with cutoff energies
proportional to the nuclear charge.  Because the model fits the CR
composition below and above the knee reasonably well, the authors
speculate on the physical grounds of the model.  While  cutoffs of the
individual components appear naturally as the result of shock
acceleration in SNRs,  Hall diffusion may dominate at high energies
facilitating particle escape from the Galaxy due to drift effects
and could provide more steepening above the knee.  The structure of
the regular magnetic field was chosen in the form proposed by Rand-Kulkarni in
the disk and incorporates a large halo with symmetric or
antisymmetric field configuration. Though the model reproduces the
knee spectrum of CR, it suffers from certain simplifications. 
Nuclear disintegration  is neglected. The transport equation includes
only diffusion and source terms with all sources located  in a ring at
4 kpc from the Galactic center. With a wider distribution of the
sources it may be difficult to explain the sharp knee structure.

Ogio \& Kakimoto (HE 1.2, p.315) suggest that the knee structure can
be explained by  particle diffusion perpendicular to the Galactic
plane due to the open magnetic field lines of loops and filaments.
However, the advection velocity derived by the authors is apparently
too large $\sim500$ km s$^{-1}$ and may conflict with that derived
from low-energy CR.

The residence time of Galactic CR in the disk has been computed by
Codino and Plouin (HE 1.2, p.319), who simulated some $10^8$
particle trajectories originating in the disk. In their simulations,
the probabilities of nuclear interactions and escape depend on the gas
column density.  The magnetic field structure adopted includes a
regular spiral field in the disk and a random component, where the
latter is responsible for chaotization of particle trajectories and
escape.  The obtained residence time depends on the nucleus charge and
has a plateau ($\sim2.2$ Myr for carbon)  between $10^{10}$ and
$10^{14}$ eV decreasing toward higher energies.  A second plateau
appears above $10^{16}$ eV with the residence time comparable to a
simple crossing time of the Galaxy.

\subsection{Galactic Sources}

Young SNRs are traditionally considered as the sites of CR
acceleration by diffusive shocks. The conventional estimates put the
maximum reachable particle energy at or just below the knee energy.
Bell \& Lucek [2] suggested that acceleration of CR particles in the
shock is accompanied by simultaneous amplification of the magnetic
field around the shock allowing for acceleration of particles to much
higher energies.  Given a possibility to increase the Alfv\'en Mach
number in the shock to $10^3$ the maximum possible particle rigidity
can reach $10^{17}$ V. This mechanism is investigated by Drury et
al. (HE 1.2, p.299) in a simplified ``box model.'' Assuming Bohm
diffusion and a Sedov-Taylor expansion law, the characteristic curve
in the momentum vs.\ time plane relates the final energies to the
starting times.  The authors indicate that under plausible conditions
this mechanism produces a high-energy tail with a slope $4.25+\delta$
which is slightly steeper than canonical value of 4 with a transition
to the usual spectrum in the knee region.

A growing number of observations have yielded large statistics of SNe
which come in various flavors: thermonuclear SNe Ia, core collapse
SNe Ib,c, II (with subclasses), and exhibit a wide range of
luminosities, expansion velocities, and chemical abundances. A new
class of core collapse SNe Id seems to be found that is also called Ic
``hypernovae'' to underline their peculiar high expansion velocities
and explosion energies.  The observed differences in SN shell
velocities and energetics are  translated into  different
efficiencies of particle acceleration and the corresponding distribution
of maximum reachable energies of accelerated particles.  This
diversity of characteristics of SNe and their different frequencies
are exploited by Sveshnikova (HE 1.2, p.307) to explain the knee
structure as a superposition of SNRs of different types. Although the
explosion energy distribution of SNe is not established,  it can
approximately be deduced from the absolute magnitude distribution. In
this model, the knee structure is mainly composed of two types of SNe
with type Ib,c contributing mostly below the knee and type IIn
dominating above the knee. The fraction of events responsible for the
formation of the knee is estimated at $\sim2\times10^{-4}$ yr$^{-1}$
with  a total energy of $\sim30\times10^{51}$ erg that might be
identified with hypernovae.  The predicted composition shows irregular
behavior above the knee with a trend toward dominance of heavier nuclei.
This obviously interesting hypothesis should be tested further.  Given
the enormous power of hypernova events and their low frequency, only a
few of them would determine the spectrum of CR at the knee region. The
consequences are a large anisotropy and large historical variations
of CR intensity.  An indication of four large increases of CR
intensity during the  last 150 kyr is indeed found, e.g., in studies
of abundance of cosmogenic $^{10}$Be in Antarctic ice [13], but their
interpretation and identification with astrophysical objects  are yet
to be done.

The sharpness of the knee and the change in the anisotropy amplitude may
be an argument toward a recent local SNR source, which adds a small
component, the knee, on the top of a smooth Galactic CR
spectrum. The explanation was originally proposed in a series of
papers by Erlykin \& Wolfendale (e.g., see [6]).  Bhadra (HE 1.2,
p.303) explores the possibility of a  pulsar origin of the knee in a
similar model. The author argues that the source of energetic
particles nearby, if it exists, should be visible in $\gamma$-rays.  A
pulsar, such as Geminga or Vela, may thus be responsible for the fine
knee structure, while the absence of the $\gamma$-rays from a SNR
nearby can be explained by the low density environment of the source.

\section{Miscellaneous}

The miscellaneous section collects several papers, which are somewhat
outside  the traditional topics of CR acceleration and propagation.
Kuwabara et al.\ (OG 1.3, p.1965) use 2-dimensional MHD simulations to
study the effect of the Parker instability in the presence of CR.  A
Lagrangian formalism for the Fokker-Planck transport equation is
proposed by Burgoa (OG 1.4, p.2011).  Zenitani \& Hoshino (OG 1.4,
p.2043) discuss a magnetic reconnection and drift kink instability in
electron/positron plasma as a source of nonthermal particles.  Saito
et al.\ (HE 1.2, p.311) investigate particle acceleration due to
electrostatic shock wave driven by counterstreaming pair plasma with
background magnetic field.

\section{Wider Perspective}

In the summary, it is appropriate to identify several
topics expect to become the subject of intensive studies in coming
years.  At low energies, ACE, Ulysses, and Voyager continue to deliver
excellent quality spectral and isotopic data, while there is no
matching experiment at GeV energies. The most accurate data in the GeV
range so far were obtained more than 20 years ago by the HEAO-3
instrument. Two generations of theorists have speculated on the origin of
the sharp peak in the B/C ratio, one of the HEAO-3 results, eagerly
awaiting new data.  New accurate positron and especially antiproton
measurements are desirable. Antiprotons with their unique spectral
shape are seen as a key to many problems, such as Galactic CR
propagation, possible imprints of our local environment, heliospheric
modulation, dark matter etc.  Happily, several high resolution space
and balloon experiments are to be launched in the near
future. PAMELA (launch in 2004) is designed to measure antiprotons,
positrons, electrons, and isotopes H through C over the energy range
of 0.1 to 300 GeV. Future Antarctic flights of a new BESS-Polar
instrument will considerably increase the accuracy of data on antiprotons
and light elements. AMS will measure CR particles and nuclei
$Z\hbox{\rlap{\hbox{\lower3pt\hbox{$\sim$}}}\lower-2pt\hbox{$<$}}26$
from GeV to TeV energies. This is complemented by measurements of
heavier nuclei $Z>29$ by TIGER.  Several missions are planned to
target specifically the high energy electron spectrum, which could
provide  unique information about our local environment and sources
of CR nearby.  They will also give necessary clues to the interpretation
of the spectrum of diffuse $\gamma$-ray emission from the future GLAST
mission capable of measuring $\gamma$-rays in the range 20
MeV -- 300 GeV.  In its turn, modeling the spectrum and distribution
of the Galactic diffuse $\gamma$-ray emission gathered by GLAST will
provide insights into the CR spectra in remote locations. Besides,
GLAST with its high sensitivity and resolution should deliver a final
proof of proton acceleration in SNRs -- long awaited by the CR
community. In the knee region and beyond, several satellite and 
ground-based
instruments are planned to dramatically increase statistics and energy
coverage (see rapporteur talks [1,17]).  A breakthrough on
supersymmetry and high-energy particle interactions should come with
operation of the new CERN large hadronic collider, LHC.

In the past and at present, new studies and discoveries in CR physics
provide a fertile ground for research in many areas of Astrophysics,
Particle Physics, and Cosmology, such as the search for dark matter
signatures, new particles and exotic physics, the origin of elements, the
origins of Galactic and extragalactic $\gamma$-ray  diffuse emission,
heliospheric modulation etc.  I am glad that this conference has been
successful enough to gather new ideas and approaches and to show
further horizons to the CR community.

\section{Acknowledgments}

I am grateful to the Organizing Committee for the invitation to be a
rapporteur and for providing the financial support as well as for their thorough preparation
of the conference.  I would like to thank all the contributing authors
for their patient  answers to my numerous questions. I am also
indebted to Vladimir Ptuskin, Frank Jones, and Andrew Strong
for discussions of various aspects of CR
acceleration and propagation.  An essential part of this report was
prepared during a visit to the Max-Planck-Institut f\"ur
extraterrestrische Physik in Garching; the warm
hospitality and financial support of the Gamma Ray Group is 
gratefully acknowledged.  This
work was supported in part by a NASA Astrophysics Theory Program grant.

\section{References}

\vspace{1\baselineskip}

\re
1.\ Battiston R.\ 2003, this conference, rapporteur talk

\re
2.\ Bell A.R., Lucek, S.G.\ 2001, MNRAS 321, 433

\re
3.\ Berezinskii V.S., Bulanov S.V., Dogiel V.A., Ginzburg V.L., Ptuskin V.S.\
1990, Astrophysics of Cosmic Rays (North Holland: Amsterdam)

\re
4.\ Bykov A.M.\ 2001, Space Science Rev.\ 99, 317

\re
5.\ Donato F.\ et al.\ 2001, ApJ 563, 172

\re
6.\ Erlykin A.D., Wolfendale A.W.\ 1997, Astropart.\ Phys.\ 7, 203

\re
7.\ Gleeson L.J., Axford W.I.\ 1968, ApJ 154, 1011

\re
8.\ Han J.L.\ 2003, Acta Astron.\ Sinica Suppl.\ 44, 148

\re
9.\ Heinbach U., Simon M.\ 1995, ApJ 441, 209

\re
10.\ Hunter S.D.\ et al.\ 1997, ApJ 481, 205

\re
11.\ Jones F.C., Ellison D.C.\ 1991, Space Science Rev.\ 58, 259

\re
12.\ Jungman G., Kamionkowski M., Griest K.\ 1996, Phys.\ Reports 267, 195

\re
13.\ Konstantinov A.N., Kocharov G.E., Levchenko V.A.\ 1990, Sov.\ Astron.\ Lett.\ 16, 343

\re
14.\ Molnar A., Simon, M.\ 2001, in Proc.\ 27th ICRC (Hamburg), 1877

\re
15.\ Moskalenko I.V., Mashnik S.G., Strong A.W.\ 2001, in Proc.\ 27th ICRC (Hamburg), 1836

\re
16.\ Moskalenko I.V., Strong A.W., Ormes J.F., Potgieter M.S.\ 2002, ApJ 565, 280

\re
17.\ Olinto A.\ 2003, this conference, rapporteur talk

\re
18.\ Parker E.N.\ 1965, Planet.\ Space Sci.\ 13, 9

\re
19.\ Roger R.S., Costain C.H., Landecker T.L., Swerdlyk C.M.\ 1999, A\&Ap Suppl.\ 137, 7

\re
20.\ Seo E.S., Ptuskin V.S.\ 1994, ApJ 431, 705

\re
21.\ Sfeir D.M., Lallement R., Crifo F., Welsh B.Y.\ 1999, A\&Ap 346, 785

\re
22.\ Sreekumar P.\ et al.\ 1998, ApJ 494, 523

\re
23.\ Takita M.\ 2003, this conference, rapporteur talk

\re
24.\ Yanasak N.E.\ et al.\ 2001, ApJ 563, 768

\endofpaper
\end{document}